\documentclass[prd, twocolumn, superscriptaddress, nofootinbib]{revtex4-1}

\usepackage{amsmath,amsfonts,amssymb}
\usepackage{graphicx}

\begin{document}

\title{Thermal and spectral dimension of (generalized) Snyder noncommutative spacetimes}
\author{Giovanni Amelino-Camelia}
\email{amelino@na.infn.it}
\affiliation{Dipartimento di Fisica Ettore Pancini, Universit\`a di Napoli "Federico II'',
and INFN, Sezione di Napoli,\\ Complesso Univ. Monte S. Angelo, I-80126 Napoli, Italy}

\author{Flaminia Giacomini}
\email{flaminia.giacomini@univie.ac.at}
\affiliation{Vienna Center for Quantum Science and Technology, Faculty of Physics,
University of Vienna, A-1090 Vienna, Austria}
\affiliation{Institute for Quantum Optics and Quantum Information,
Austrian Academy of Sciences, A-1090 Vienna, Austria}

\author{Giulia Gubitosi}
\email{g.gubitosi@science.ru.nl}
\affiliation{Radboud University, Institute for Mathematics, Astrophysics and Particle Physics, Heyendaalseweg 135, NL-6525 AJ Nijmegen, The Netherlands}\affiliation{Dipartimento di Fisica, Universit\`a di Roma ``La Sapienza'', P.le A. Moro 2, 00185 Roma, Italy}

\begin{abstract}
We report an investigation of the Snyder noncommutative spacetime and of some of its most natural generalizations, also looking at them
as a powerful tool for comparing different notions of dimensionality of a quantum spacetime.
It is known that (generalized-)Snyder noncommutativity, while having rich off-shell implications (kinematical Hilbert space),
does not affect on-shell particles (physical Hilbert space), and we argue that physically meaningful notions of dimensionality
 should describe such spacetimes as trivially four-dimensional, without any running with scales.
By studying the thermodynamics of a gas of massless particles living on these spacetimes, we find that indeed the Snyder model and its generalizations have constant thermal dimension of four.  We also compute the spectral dimension of the Snyder model and its generalizations, finding that, as a result of its sensitivity to off-shell properties, it runs from the standard value of four in the infrared towards lower values in the ultraviolet limit.
\end{abstract}

\maketitle

\section{Introduction}

The Snyder noncommutative spacetime \cite{Snyder:1946qz} is the earliest and one of the most studied proposals for describing quantum properties of spacetime at the Planck scale.
It is generally understood as an interesting case of spacetime whose coordinates have discrete spectrum\cite{Lu:2011it,Amelino-Camelia:2014mea},
as one can most rigorously establish within a
manifestly-covariant analysis \cite{Rovelli:1989jn, Reisenberger:2001pk, Gambini:2000ht, Halliwell:2002th} on the kinematical Hilbert space.
However, this discreteness leaves no trace\footnote{Similar conclusions were also reached in studies \cite{Mignemi:2013aua} of the semiclassical  limit of the Snyder model.} on the physical Hilbert space \cite{Amelino-Camelia:2014mea}, obtained by enforcing
the on-shellness requirement for physical particles. In an appropriate sense it is a spacetime whose unphysical/abstract points are quantized but whose physical events (crossing of worldlines of on-shell particles) are indistinguishable from those of ordinary (commutative)  Minkowski spacetime.
Here we provide further results in support of this picture
 by analyzing the thermodynamical properties\footnote{Some thermodynamical implications of  Snyder noncommutativity
 were previously studied in \cite{Nozari:2015iba}, but only considering
a ``nonrelativistic" (Galileian) regime and restricting the noncommutativity to three-dimensional Euclidean space.}
 of a gas of photons living on Snyder spacetime. We also show that the same conclusions are reached in
 its generalization introduced in \cite{Quesne:2006fs,Quesne:2006is} as a covariant extension of the deformed model of Heisenberg phase space proposed in \cite{Kempf:1996fz}.

 Our main objective, however, is to
 propose the Snyder model and its generalizations as a rather natural arena for
 testing
 different proposals for the notion of dimensionality of a quantum spacetime:
 in light of the observations we just summarized,
physically meaningful notions of dimensionality
 should describe such spacetimes as trivially four-dimensional, without any running with scales.
 
In order to make our case for this, we use the (generalized) Snyder model to compare the behavior of two different notions of dimensionality that can be applied to quantum spacetime. The thermal dimension, first proposed in \cite{Amelino-Camelia:2016sru}, is computed by exploiting the fact that some thermodynamical properties of a gas of massless particles living on standard Minkowski spacetime scale with temperature in a way that depends on the dimensionality of the spacetime itself. One can then associate an effective dimension to the quantum spacetime by  looking at the scaling with temperature  of the same quantities, evaluated for a gas of massless particles living on it. The spectral dimension is a geometrical notion of dimensionality that has been used in several approaches to quantum gravity research \cite{Ambjorn:2005db, Benedetti:2009ge, Benedetti:2008gu, Calcagni:2013vsa, Litim:2003vp, Horava:2009if, Modesto:2008jz, Amelino-Camelia:2013cfa, Arzano:2014jfa}. It is deeply linked to the properties of the Euclideanized d'Alembertian of the theory, and can be thought of as the effective dimension probed by a fictitious diffusion process on the Euclideanized spacetime.

In the following Section \ref{sec:SnyderDef} we present the Snyder model and its generalizations, defined by the deformed Heisenberg commutators of phase space coordinates. We  construct the kinematical and physical Hilbert spaces, as first developed in \cite{Amelino-Camelia:2014mea} for the Snyder case. We show that the nontrivial spacetime properties encoded in the kinematical Hilbert space (\emph{e.g.} coordinate discreteness) do not survive in the restriction to the physical Hilbert space, that behaves as that of undeformed Minkowski spacetime. To make this point stronger, and emphasize that it is a significant result, in Subsection \ref{sub:kMHS} we contrast it to the nontriviality of both  the kinematical and physical Hilbert spaces of the $\kappa$-Minkowski model for spacetime noncommutativity \cite{majid1994bicrossproduct, lukierski1995classical}.  The fact that the (generalized) Snyder geometry is discrete, but this discreteness is not observable, leads us to argue  that indeed the model is a good test ground for evaluating the sensitivity to unphysical spacetime properties of different notions of dimensionality.   In particular, we expect that any physically relevant notion of dimensionality should not be affected by the unobservable  discreteness characterizing the (generalized) Snyder spacetime. In Section \ref{sec:thermaldimension} we show that indeed the thermal dimension is not sensitive to unphysical properties, since it is constantly equal to the standard value of four for both the Snyder model and its generalizations. As a cross-check, we show that however the thermal dimension does show a running if we modify the physical properties of the model by taking higher-order functions of the energy-momentum invariant to characterize the dispersion relation. So the thermal dimension does flag physically relevant departures from standard Minkowski spacetime. The spectral dimension is dealt with in Section \ref{sec:spectraldimension} and it demonstrates to be dominated by the unphysical geometrical properties of the (generalized) Snyder spacetime, with its value running from the standard value of four in the infrared to smaller values in the ultraviolet. Matters become even worse if we modify the on-shell condition as done for the thermal dimension. In fact, in this case the spectral dimension it not always well defined, because of divergences introduced by the Euclideanized d'Alembertian.

\section{The Snyder model and its generalizations}
\label{sec:SnyderDef}

The Snyder model is characterized by the following deformed Heisenberg relations between phase space coordinates:
\begin{eqnarray}
	 		\left[x^\mu,\, p^\nu\right]&=& -i  \left[ \eta^{\mu \nu}-\lambda^{2} p^\mu p^\nu \right],\nonumber\\
	 		\left[x^\mu,\, x^\nu\right]&=&-i \lambda^{2}\left(p^\mu x^\nu - p^\nu x^\mu \right), \label{eq:Snyder}\\
	 			\left[p^\mu,\,p^\nu\right]&=&0,\nonumber
	 \end{eqnarray}
where $\lambda$ is the length scale of coordinates noncommutativity, $\eta^{\mu\nu}=\text{diag}[1,-1,-1,-1]$ and we adopted units such that $c=1$ and $\hbar=1$.
A generalization of this was proposed in \cite{Quesne:2006fs,Quesne:2006is}, that introduced further corrections to the commutators, depending on a new length scale $\bar\lambda$, in principle independent from $\lambda$:
\begin{eqnarray}
	 		\left[x^\mu,\, p^\nu\right]&=&-i  \left[ (1-\bar\lambda^2 p_\rho p^\rho)\eta^{\mu \nu}-\lambda^{2} p^\mu p^\nu \right],\nonumber\\
	 		\left[x^\mu,\, x^\nu\right]&=&i  \frac{2\bar\lambda^2-\lambda^2-\bar\lambda^2(2\bar\lambda^2+\lambda^2)p_\rho p^\rho}{1-\bar\lambda^2 p_\rho p^\rho}\left(p^\mu x^\nu - p^\nu x^\mu \right),\nonumber\\
	 		\left[p^\mu,\,p^\nu\right]&=&0,\label{eq:genSnyder}
	 \end{eqnarray}
such that the Snyder model is recovered for $\bar\lambda=0$. 

While the Snyder commutation rules in Eq.~\eqref{eq:Snyder} are compatible with standard generators of relativistic symmetries,  for the  more general model in Eq.~\eqref{eq:genSnyder}  a deformed representation on phase space of rotation generators $\mathcal{L}_{ij}$ and of boost generators  $\mathcal{L}_{i0}$ is required for  compatibility  with the symplectic structure of Eq.~\eqref{eq:genSnyder} and with undeformed commutators of the Lorentz sector of the Poincar\'e algebra:
	 \begin{equation}
			\mathcal{L}_{\mu\nu}= \left( 1-\bar\lambda^2 p_\rho p^\rho \right)^{-1} \left( x_\mu p_\nu - x_\nu p_\mu \right).\label{eq:Lorentz}
		\end{equation}
Given that for both the Snyder model and its generalization the algebra of  symmetry generators is undeformed, in both cases the Casimir takes the usual special-relativistic form,
\begin{equation}
\mathcal C= p_{\mu}p^{\mu}\,.\label{eq:Casimir}
\end{equation}

The construction of the kinematical Hilbert space of the (generalized) Snyder model relies on the representation of the noncommuting coordinates $x^{\mu}$ on the operators $\{q^{\mu},p^{\mu}\}$ that satisfy standard Heisenberg relations,
\begin{align}
	\left[ {q}^{\mu}, {q}^{\nu} \right]&=\left[ {p}^{\mu}, {p}^{\nu} \right]= 0, \nonumber \\
	\left[ {q}^{\mu}, {p}^{\nu} \right]&=-i\eta^{\mu\nu}\,, \label{eq:pregeometry}
\end{align}
and are the phase-space operators of  covariant quantum mechanics \cite{Rovelli:1989jn, Reisenberger:2001pk, Gambini:2000ht, Halliwell:2002th}. One can easily find a family of viable representations,
\begin{equation}
x^{\mu}=(1- \bar\lambda^{2} p_{\nu}p^{\nu}) q^{\mu}-\lambda^{2}p^{\mu}p^{\nu}q_{\nu}+ i \gamma p^{\mu},
\end{equation}
spanned by the new parameter $\gamma$, with dimensions of a length squared.\footnote{Note that this parameter has no role in the definition of the model itself, since it does not enter in the  commutators of the phase space coordinates, Eq.~\eqref{eq:genSnyder}, and is only relevant in the mapping between the noncommuting coordinates and the  auxiliary operators $\{q^{\mu},p^{\nu}\}$.}
This representation reduces to the one found in \cite{Amelino-Camelia:2014mea} for the Snyder model in the limit $\bar\lambda=\gamma=0$. The states of the kinematical Hilbert space are normalizable wave functions, whose scalar product  can be written in the momentum space representation as 
\begin{equation} 
	\langle \psi|\phi\rangle = \int d\mu(p) \psi^* (p)  \phi(p)\,.
\end{equation}
The measure $d\mu(p)$ is defined by the requirements that the coordinates $x^{\mu}$ be Hermitian operators on the kinematical Hilbert space (so that the kinematical Hilbert space actually encodes the geometrical properties of the noncommutative spacetime) and that the scalar product is invariant under the action of the Lorentz generators in Eq.~\eqref{eq:Lorentz}. 
These conditions select the following measure \cite{Quesne:2006is}:
\begin{equation}
d \mu (p) =  d^{4}p \left[1-(\bar\lambda^2+\lambda^2)p^\mu p_\mu\right]^{-\alpha}\,, \label{eq:dmu}
\end{equation}
with $\alpha= \frac{2\bar\lambda^2+5\lambda^2-2\gamma}{2(\bar\lambda^2+\lambda^2)}$.
Because of this deformed measure, the kinematical Hilbert space of the (generalized) Snyder model is nontrivial and correspondingly encodes nontrivial properties of the spacetime coordinates $x^{\mu}$, such as a discrete spectrum \cite{Amelino-Camelia:2014mea}. However, physical observables are not defined on the kinematical Hilbert space, but on the physical one, which is the restriction of the Hilbert space to states that satisfy the Hamiltonian constraint $\mathcal C-m^{2}=0$. And, as we  show in the following, in a non-interacting theory (generalized) Snyder noncommutativity leaves no trace on the physical Hilbert space.  

The physical constraint (along with the restriction to positive-energy states) can be enforced at the level of the scalar product \cite{Amelino-Camelia:2014mea}:
\begin{equation}\label{eq:physHilbSp}
\langle \psi|\phi\rangle_{\text{phys}}=\int  d\mu (p) \delta(p^{\mu}p_{\mu}-m^{2})\Theta(p_{0})\psi^{*}(p)\phi(p)\,.
\end{equation}
Once on-shell, the deformed integration measure \eqref{eq:dmu} reduces to the trivial measure multiplied by a function of the mass, which  can be reabsorbed in the state normalization. Then, after integration over the energy, the physical scalar product takes the standard form,
\begin{equation}
\langle \psi|\phi\rangle_{\text{phys}}\sim \int  \frac{d^{3}\vec p}{2 \omega_{\vec p}} \psi^{*}(\omega_{\vec p},\vec p)\phi(\omega_{\vec p},\vec p)\,,
\end{equation}
where $\omega_{\vec p}=\sqrt{|\vec p|^{2}+m^{2}}$. And not only is the scalar product undeformed, but also any observable on the physical Hilbert space is so. \footnote{This was demonstrated in \cite{Amelino-Camelia:2014mea} for the Snyder model. The proof can be easily  extended to the generalized model of Eq.~\eqref{eq:genSnyder} by noting that the modification of the Lorentz generators described in Eq.~\eqref{eq:Lorentz} reduces to a constant multiplicative factor once the on-shellness constraint is enforced.} This brings us to the conclusion we anticipated in the Introduction, that while the nontrivial properties of (generalized) Snyder spacetime are relevant in the unphysical kinematical Hilbert space, they leave no trace on the physical Hilbert space (of free particles). For this reason the (generalized) Snyder model is an ideal test ground for proposals to characterize the dimensionality of quantum spacetime. In fact, the results just exposed motivate the requirement that any physically relevant notion of dimensionality should also be insensitive to the coordinate discreteness, and in particular it should evaluate the dimension to be constantly equal to the standard value of four.

Before proceeding to analyze the behavior of different  notions of dimension applying them to the (generalized) Snyder model, we find it useful to pause and compare the results on the kinematical and physical Hilbert spaces of the (generalized) Snyder model to the corresponding ones valid for another much studied model of spacetime noncommutativity, known as $\kappa$-Minkowski \cite{majid1994bicrossproduct, lukierski1995classical}. This comparison will turn out to be very insightful, since we will show that the triviality of the physical Hilbert space is peculiar to the (generalized) Snyder model, while in $\kappa$-Minkowski the traces of coordinates noncommutativity are not lost in going from the purely geometrical description of spacetime provided by the kinematical Hilbert space to the physical description encoded in the Hilbert space. For this reason, the $\kappa$-Minkowski model would not be a good case study for testing the physical relevance of different notions of dimensionality, since one would not be able to discern whether any nontrivial property that is eventually found is to  be ascribed to the  physically relevant on-shell theory or is just an unphysical artifact. So it is exactly the triviality of the physical Hilbert space of the (generalized) Snyder model
that makes it a good candidate to evaluate the meaningfulness of different proposal for a notion of quantum spacetime dimension.

\subsection{Aside: kinematical vs. physical Hilbert space of $\kappa$-Minkowski noncommutative spacetime}
\label{sub:kMHS}

The $\kappa$-Minkowski  spacetime is characterized by the following commutators between spacetime coordinates \cite{majid1994bicrossproduct, lukierski1995classical}:
\begin{equation}
	\left[ {x}_i, {x}_0 \right] = \frac{i}{\kappa} {x}_i; \qquad \left[ {x}_i, {x}_j \right] = 0,
\end{equation}
where $i=\{1,2,3\}$ and $ \kappa^{-1}$ is a length scale. The construction of the kinematical and physical Hilbert spaces can be performed in a completely similar way as discussed above. The representation of  the noncommuting coordinates on the phase space operators of Eq.~\eqref{eq:pregeometry} reads \cite{amelino2013relative, Amelino-Camelia:2013nza}:
\begin{equation}
	{x}_0 = {q}_0, \qquad {x}_i = {q}_i e^{ {p}_0/\kappa}.
\end{equation}
The relativistic symmetries of $\kappa$-Minkowski are deformed, and are described by the $\kappa$-Poincar\'e algebra \cite{majid1994bicrossproduct, lukierski1995classical}. The action of the symmetry generators on the noncommutative spacetime in terms of the phase space coordinates \eqref{eq:pregeometry} was given in \cite{amelino2013relative}.  For our purposes it is sufficient to note that 
the Casimir of the $\kappa$-Poincar\'e algebra can be represented as
\begin{equation}
	\mathcal{C}_{\kappa} = \left(2 \kappa\right)^2 \sinh^2\left( \frac{ {p}_0}{2\kappa} \right) - e^{- {p}_{0}/\kappa} |\vec{p}|^2,
\end{equation}
and is used to define the on-shell constraint for point particles of mass $m$:
\begin{equation}
\mathcal{C}_{\kappa}-m^2=0\,.\label{eq:HamiltonContraint}
\end{equation}

The kinematical Hilbert space is again equipped with the scalar product 
\begin{equation} 
	\langle \psi|\phi\rangle = \int d\mu(p) \psi^* (p)  \phi(p)\,,
\end{equation}
where now the measure on momentum space is of course required  to be invariant under the $\kappa$-Poincar\'e relativistic symmetries: 
\begin{equation}
	d\mu(p)= d p_{0} d^{3} \vec p e^{-3 p_0/\kappa}.\label{eq:kMmeasure}
\end{equation}
The physical Hilbert space, obtained after enforcing the Hamiltonian constraint \eqref{eq:HamiltonContraint}, is characterized by the scalar product
\begin{equation} 
	\langle \psi|\phi\rangle_{\text{phys}} = \int d\mu(p) \delta(\mathcal{C}_\kappa - m^2)\Theta(p_0)\, \psi^* (p)  \phi(p)\,.\label{eq:physHSkM}
\end{equation}
In contrast to what happens in the (generalized) Snyder case, where the fact that the integration measure \eqref{eq:dmu} only depends on momenta via the Casimir turned out to be crucial in establishing the triviality of the physical Hilbert space, it is easy to convince oneself that the scalar product  of Eq.~\eqref{eq:physHSkM} is still nontrivial. In fact, not only does the momentum space measure  \eqref{eq:kMmeasure}  not reduce to the standard one when evaluated on-shell, but the  Hamiltonian constraint is itself deformed. 
For this reason, the fuzziness characterizing $\kappa$-Minkowski coordinates \cite{amelino2013relative} leaves observable traces in the fuzziness of trajectories of freely-propagating particles \cite{Amelino-Camelia:2013nza}.
One would then expect that any notion of distance that is sensitive to physically relevant properties of  spacetime would give nontrivial result when applied to the $\kappa$-Minkowski models. However, it would do so just as a notion of distance that is  sensitive to off-shell properties would. For this reason, models such as the $\kappa$-Minkowski one are not ideally suited to select physical notions of dimensionality. 

\section{Thermal Dimension of the (generalized) Snyder model}
\label{sec:thermaldimension}

Having motivated the appropriateness of the (generalized) Snyder model in testing the physical relevance of different notions of dimensionality, in this Section we analyze the behavior of the thermal dimension. 

In a  gas of  standard special-relativistic massless particles, both the Stefan-Boltzmann law, describing the scaling of the total energy of the gas $U$ with its temperature $T$, and the equation of state parameter,  relating the energy density $\rho$ and pressure $P$, depend on the dimension of spacetime in a well-defined way. Specifically, if the gas of photons lives in a $D+1$ dimensional spacetime, then the Stefan-Boltzmann law reads
\begin{equation}
U\propto T^{D+1}\,,\label{eq:SB}
\end{equation}
while the equation of state parameter $w\equiv P/\rho$ equals
\begin{equation}
w=\frac{1}{D}\,.\label{eq:EOS}
\end{equation}
The thermal dimension of a quantum spacetime can be computed by considering a gas of photons living on such a space and comparing the  behavior of the Stefan-Boltzmann law and of the equation of state to those described by Eqs.~\eqref{eq:SB}-\eqref{eq:EOS}. In this way, one can associate an effective dimension to the quantum spacetime model \cite{Amelino-Camelia:2016sru}. In the following, we will focus on a gas of photons whose phase space and symmetries are described by the (generalized) Snyder model defined in the previous Section.

The thermodynamics of a gas of massless particles is encoded in the partition function. The relevant possibly nontrivial  ingredients that contribute to its evaluation are the on-shell relation and the measure of integration on momentum space \cite{Amelino-Camelia:2016sru, Gorji:2016gfr}. As was shown in the previous Section, for the (generalized) Snyder model only the integration measure is deformed (see Eq. \eqref{eq:dmu}), while the on-shellness is codified by the standard Casimir, Eq. \eqref{eq:Casimir}. So the  associated partition function can be written in a covariant form as:
\begin{eqnarray}
	\log Q &=& -\frac{V}{(2\pi)^3} \int d\mu (p)\Big[ \delta(\Omega)\Theta (p^{0})\cdot\nonumber\\ && \cdot2p^{0} \,\log \left( 1 - \text{e}^{- p^{0}/k_{B}T} \right)\Big]\,,
\end{eqnarray}
where $k_B$ is the Boltzmann constant and $\delta (\Omega)$ enforces the on-shell relation $\mathcal C=0$.  We can see already at this level that there is no reason to expect the thermal dimension to behave nontrivially. In fact, the only deformation with respect to the standard case resides in  the momentum space measure, that however enters in the picture in the same way as it did in the definition of the physical Hilbert space, Eq.~\eqref{eq:physHilbSp}.  And, as discussed in that context,  once the   on-shell condition is enforced on the measure, its nontrivial contribution becomes an irrelevant constant factor. This is in fact what happens also in the partition function.
After integrating out the energy and enforcing the on-shell relation, the partition function reads
\begin{equation}
	\log Q\propto -\frac{V}{(2\pi)^3} \int d^3 \vec p \log \left( 1 - \text{e}^{-|\vec p|/k_B T} \right),
\end{equation}
which is the same expression as for a standard gas of photons. So  the thermodynamical properties of (generalized) Snyder photons are undeformed and the thermal dimension is constant and equal to four. We would like to emphasize that the fact that the deformed momentum space measure is trivial on-shell played a crucial role in reaching this result.

\begin{figure}[h]
\centering
\scalebox{0.65}{\includegraphics{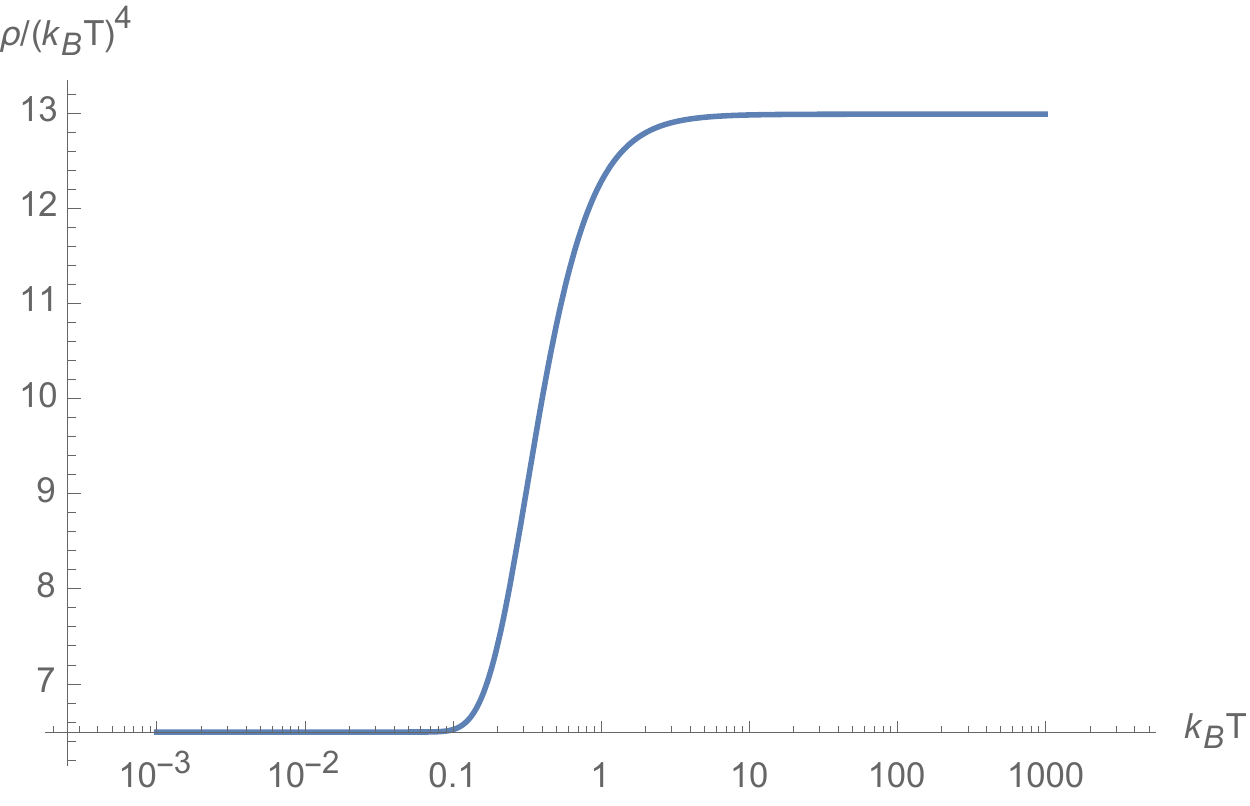}}
\caption{\label{fig:SB1} Temperature dependence of the  energy density of a gas of photons  whose partition function is given by Eq. \eqref{eq:OmegaXi}, with $\xi=1$. The temperature is in Planckian units, while the energy density is in arbitrary units.  }
\end{figure}

\begin{figure}[h]
\centering
\scalebox{0.65}{\includegraphics{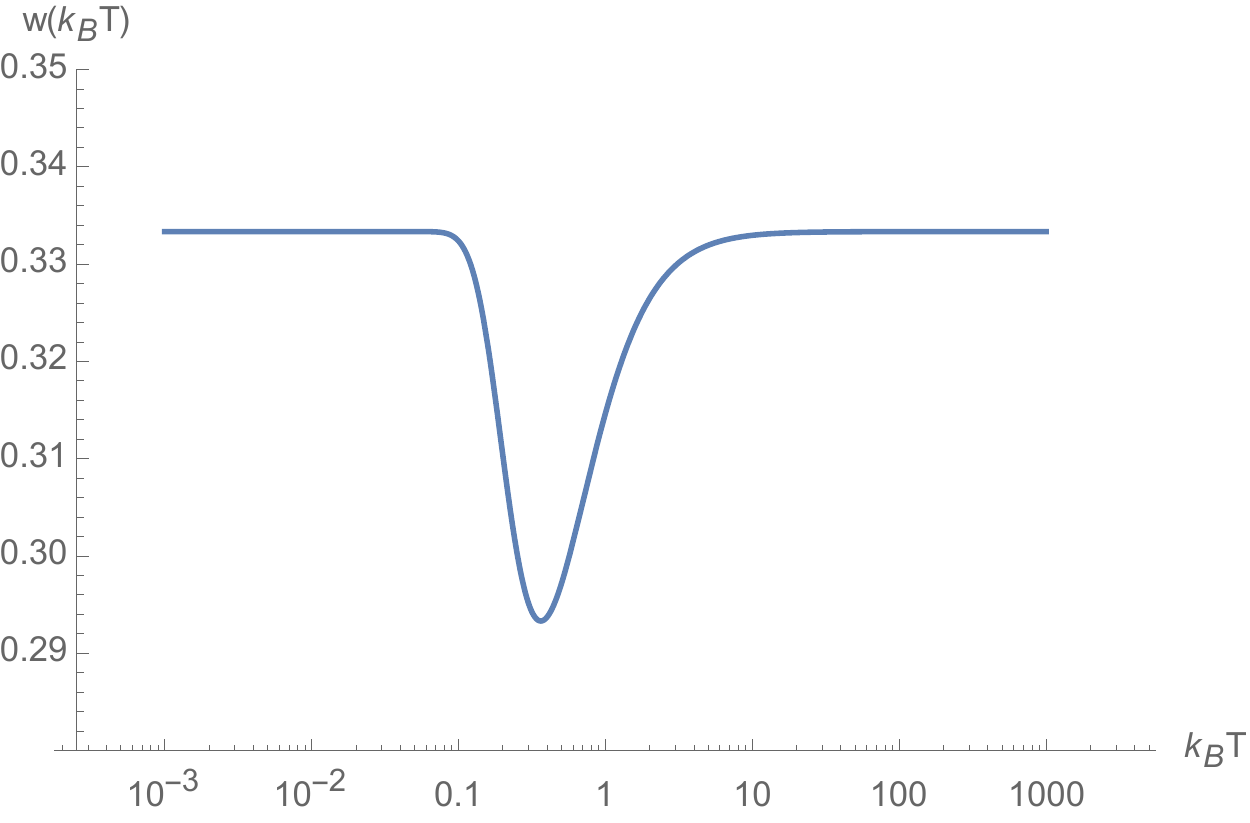}}
\caption{\label{fig:SB2} Temperature dependence of the  equation of state parameter of a gas of photons  whose partition function is given  by Eq. \eqref{eq:OmegaXi}, with $\xi= 1$. The temperature is in Planckian units.}
\end{figure}

\begin{figure}[h]
\centering
\scalebox{0.65}{\includegraphics{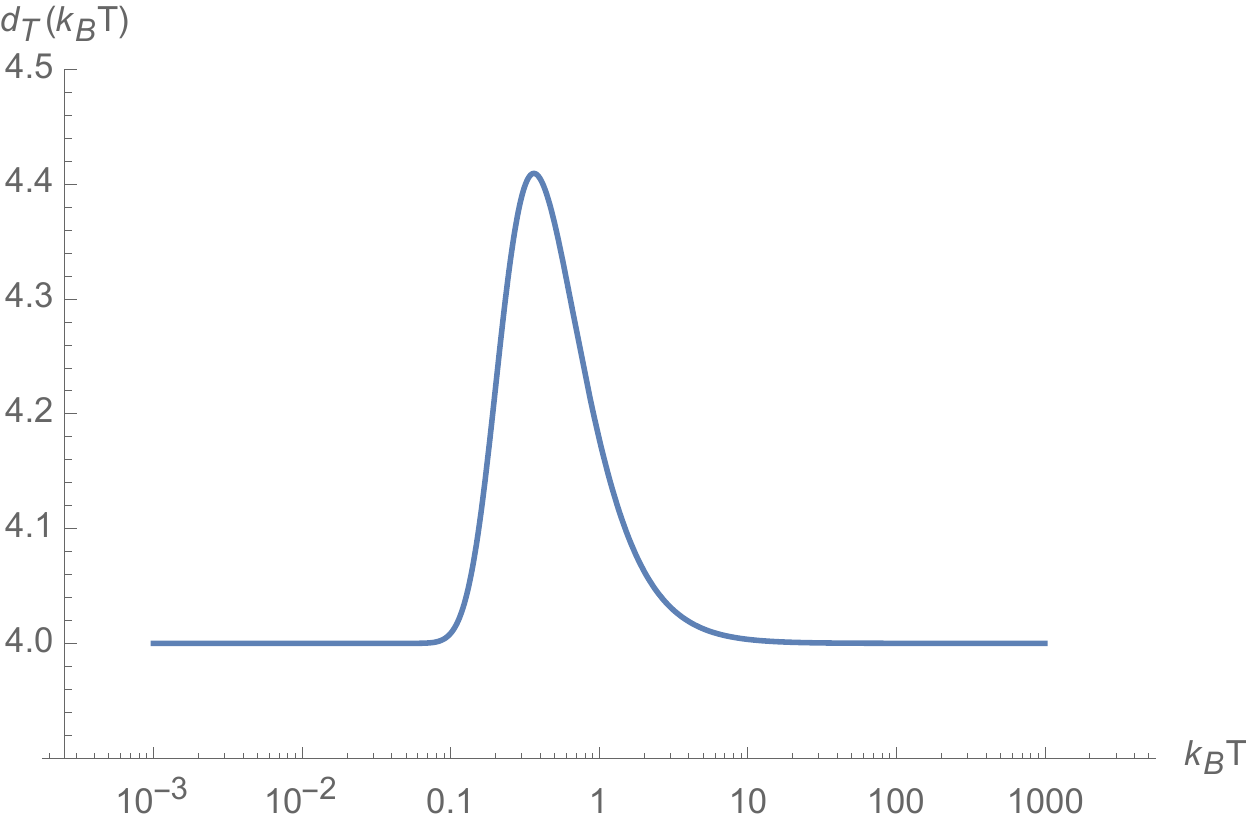}}
\caption{\label{fig:SB3} Temperature dependence of the  thermal dimension of a gas of photons  whose partition function is given  by Eq. \eqref{eq:OmegaXi}, with $\xi= 1$. The temperature is in Planckian units.}
\end{figure}

In order for the reader to compare this  case where the thermal dimension is undeformed to one where the model has physically relevant nontrivial properties, and verify that the   thermal dimension does flag these, we  modify the physical constraint of the Snyder model, using a function of the Casimir \eqref{eq:Casimir} to define the on-shellness condition: 
\begin{equation}
\Omega_\xi(p)\equiv\mathcal{C}-\ell^{2\xi}\mathcal{C}^{1+\xi} = 0, \label{eq:CasimirXi}
\end{equation}
 where $\xi$ is a positive integer and $\ell$ is a positive constant with the dimension of a length, in principle independent from the other parameters of the model. Note that, being it given by a function of the Casimir, this constraint is still compatible with the (standard) relativistic symmetries of Snyder spacetime. 
The partition function then reads:
 \begin{eqnarray}
 \log Q_{\xi} &\propto& -\frac{V}{(2\pi)^3} \int d\mu (p)\Big[ \delta(\Omega_{\xi})\Theta (p^{0})\cdot\nonumber\\
 &&\cdot2p^{0} \,\log \left( 1 - \text{e}^{- p^{0}/k_{B}T} \right)\Big]\,.\label{eq:OmegaXi}
 \end{eqnarray}
This  is completely analogous to the partition function of the Asymptotic-Safety-inspired model studied in \cite{Amelino-Camelia:2016sru}. As discussed in more detail there, the partition function receives  contributions from the two positive-energy solutions of the on-shell condition:
\begin{equation} \label{eq:delta2}
	\delta (\Omega_\xi) = \frac{\delta (p^{0}- |\vec p|)}{2|\vec p|} + \frac{\delta (p^{0}- \sqrt{|\vec p|^2 + \frac{1}{\ell^2}})}{2\xi \sqrt{|\vec p|^2 + \frac{1}{\ell^2}}}\,.
\end{equation}
At low temperatures only the first term is relevant, while at super-Planckian temperatures the two contributions are effectively equivalent. So in these two regimes  the thermal dimension takes the standard value of four. \footnote{Because of the doubling of equivalent contributions, at high temperatures the numerical value of the energy density is twice as much the value at low temperatures, but is scales in the usual way with the temperature, so it does not affect the thermal dimension.} It is only at intermediate scales that the two on-shell relations are relevantly different and both contribute to the partition function. And in fact at intermediate scales, around the Planck scale, the Stefan-Boltzmann law and the equation of state have a nontrivial behaviour, as shown in Figures \ref{fig:SB1} and \ref{fig:SB2}. Accordingly, at intermediate scales the thermal dimension takes values higher than four, as shown in Figure \ref{fig:SB3}.


\section{Spectral dimension of the (generalized) Snyder model}
\label{sec:spectraldimension}

As mentioned in the Introduction, the spectral dimension is the effective dimension probed by a fictitious diffusion process on the Euclideanized spacetime. It is determined by the dependence of the average return probability on the (fictitious) diffusion time. The average return probability related to the diffusion process is computed as
\begin{equation} \label{eq:AvRetProb}
	P(s) \propto \int d\mu^{(E)}(p) e^{ s \,\Omega^{(E)} (p)}\,,
\end{equation}
where $s$ is the diffusion time, $d\mu^{(E)}(p)$ is the Euclideanized measure on momentum space  and $\Omega^{(E)} (p)$ is the Euclidean d'Alembertian. The spectral dimension is then defined as:
\begin{equation}
	d_s(s)= -2 \frac{\partial \log P(s)}{\partial \log s}\,.
\end{equation}
The ultraviolet value of the spectral dimension, which is sensitive to the (ultra-)Planck-scale properties of the theory, is obtained in the $s\rightarrow 0$ limit.  It was already suggested \cite{Amelino-Camelia:2016sru, Carlip:2017eud}  that, while the spectral dimension could possibly be useful in characterizing some geometrical properties of quantum spacetime, is it however unphysical, being it sensitive to the off-shell modes of the theory. In the following we expose this by computing the spectral dimension of the (generalized) Snyder model, for which we have shown above that all nontrivial features are unphysical, at least at the level of the free theory.

\begin{figure}[h]
\centering
\scalebox{0.55}{\includegraphics{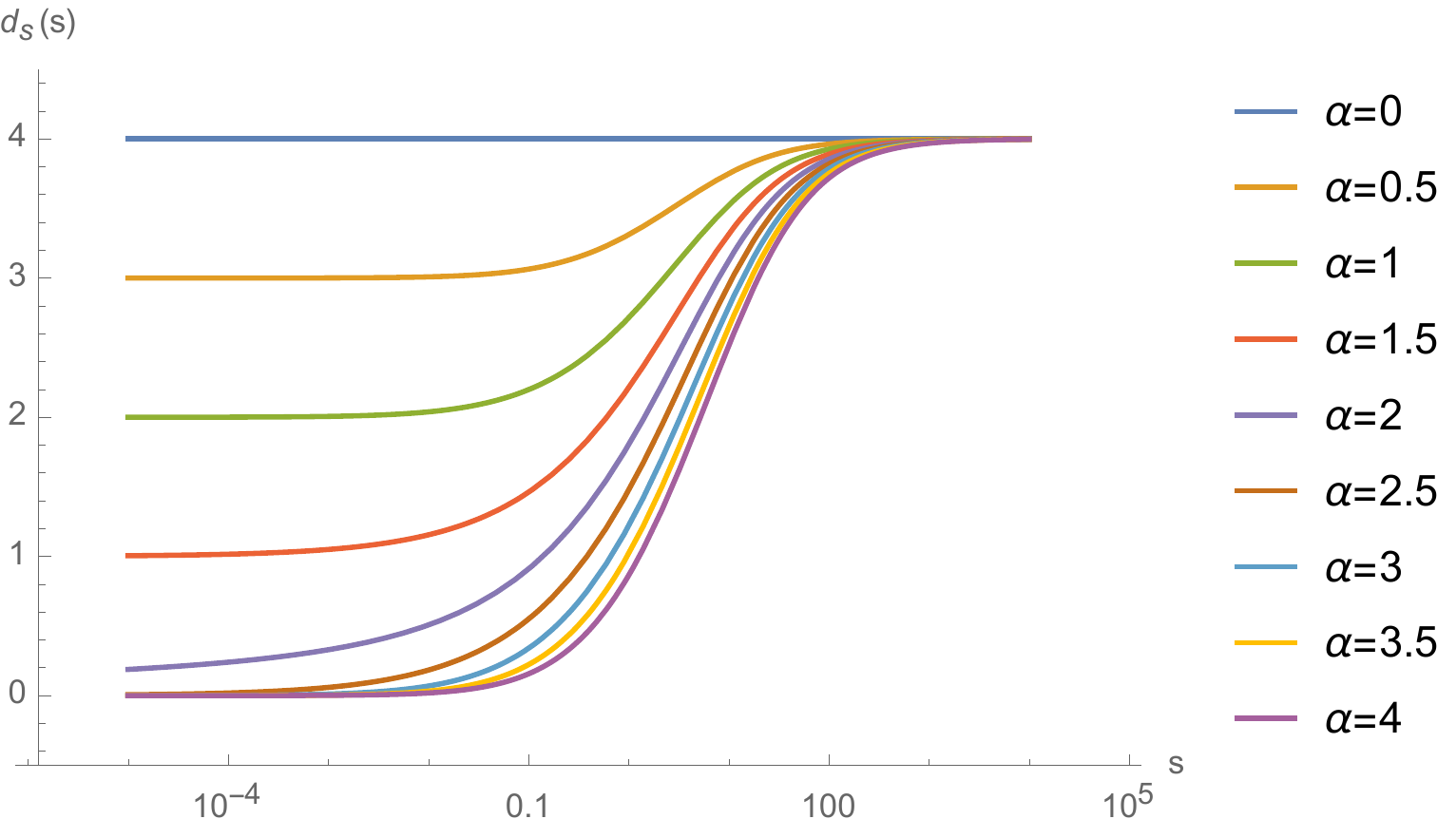}}
\caption{Spectral dimension deduced from the average return probability in Eq. \eqref{eq:Psspherical}  as a function of $s$ and for different values of $\alpha$, with $\bar\lambda^2=\lambda^2=1$.}\label{fig:Sp_dim_alpha}
\end{figure}

\begin{figure}
	\centering
	\includegraphics[scale=0.55]{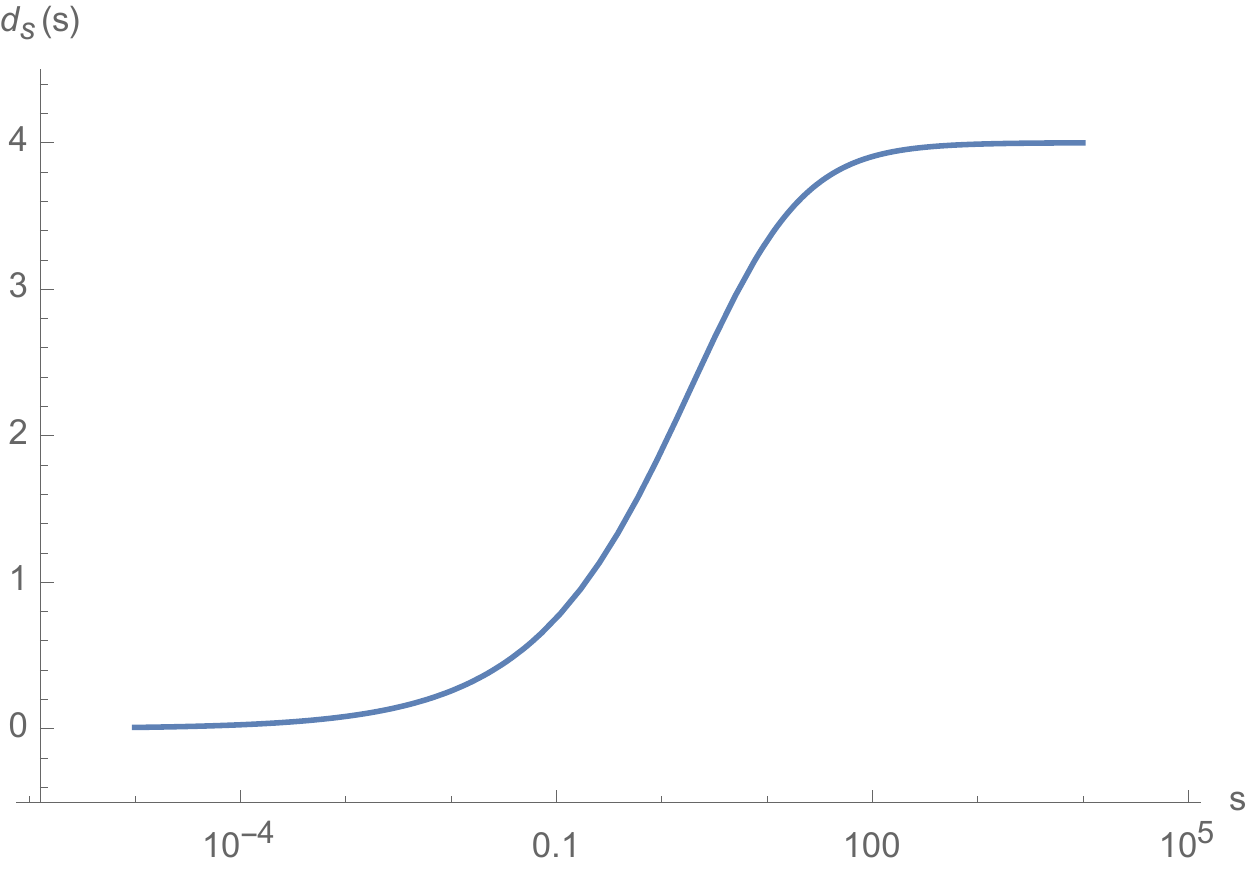}
	\caption{Running of the spectral dimension in Snyder model, as deduced from the average return probability in Eq. \eqref{eq:Psspherical} with $\bar\lambda^2=0$, $\lambda^2=1$ and $\alpha=\frac{5}{2}$.}\label{fig:Snyder_sp_dim}
\end{figure}

Using the standard choice of Casimir, Eq. \eqref{eq:Casimir}, the average return probability reads:
\begin{eqnarray}
	P(s)&\propto& \int	d^4 p \left[1+(\bar\lambda^2+\lambda^2)(p_0^2 +|\vec p|^{2})\right]^{-\alpha} e^{-s (p_{0}^{2}+|\vec p|^{2})}\nonumber\\
	& \propto& \int	d r \left[1+(\bar\lambda^2+\lambda^2)r^2\right]^{-\alpha}r^3 e^{-s r^2} \,.\label{eq:Psspherical}
\end{eqnarray}
Note that we have performed a Wick rotation $p^0 \rightarrow i p^0$ on both the integration measure and the D'Alembertian and in the second step we have introduced spherical coordinates $r^2\equiv p_0^2+ |\vec p|^{2}$ (the angular variables can be integrated out and do not contribute to the spectral dimension).
The behavior of the spectral dimension is shown in Figure \ref{fig:Sp_dim_alpha} for different values of $\alpha$, keeping $\bar\lambda^2=\lambda^2=1$. As expected, in the low-energy ($s\rightarrow \infty$) limit the spectral dimension always takes the standard value of four. In the UV regime ($s\rightarrow 0$), for any $\alpha\neq 0$ (in which case the integration measure becomes trivial) the spectral dimension runs to lower values, that tend to zero as the value of $\alpha$ increases. The value of $\alpha$ characterizing the canonical Snyder model, $\alpha=\frac{5}{2}$,  is the lowest value of $\alpha$ for which the spectral dimension goes to zero in the UV. We also plot the behavior of the spectral dimension in the standard Snyder model ($\bar\lambda^2=\gamma=0$), see Figure \ref{fig:Snyder_sp_dim}.

While the results just presented already signal that the spectral dimension is not to be trusted in characterizing the physical properties of quantum spacetime models, it is quite suggestive that we have to face  even bigger  difficulties when accounting for the possibility that the on-shell relation is given by some function of the Casimir, as in Eq. \eqref{eq:CasimirXi}. In fact, not all of the possible choices of Casimir that we have considered for the thermal dimension, Eq. \eqref{eq:CasimirXi}, admit a well-defined spectral dimension. This is not surprising, since  the spectral dimension is defined on a Euclideanized version of the model, and in general the Euclidean and Lorentzian versions of a quantum gravity model can be profoundly different \cite{Carlip:2015mra, Carlip:2017eud}.  Adopting this new on-shell condition, the Wick-rotated form of the D'Alembertian is
\begin{equation}
- (p_{0}^{2}+|\vec p|^{2}) -\ell^{2\xi} (-1)^{1+\xi}(p_{0}^{2}+|\vec p|^{2})^{1+\xi}\,,
\end{equation}
and it is immediate to see that for even values of $\xi$ the return probability diverges so that it is not possible to compute the spectral dimension.
If we restrict to odd values of $\xi$, the average return probability is
\begin{eqnarray}
	P(s)\propto \int	d r \left[1+(\bar\lambda^2+\lambda^2)r^2\right]^{-\alpha}r^3 e^{-s r^2 (1+\ell r)^{2\xi}} \,,\label{eq:PssphericalXi}
\end{eqnarray}
and one finds a running of the spectral dimension that has similar features as the ``standard Casimir case'' studied above. The role of $\xi$ is merely to further shift the UV value of the spectral dimension towards zero.

\section{Closing remarks}

Recent research in quantum gravity has  devoted quite some effort to describing the dimension of  quantum spacetime. We were here concerned with the identification of notions of dimensionality that are informative with respect to the physical content of the theory.

We argued  that the Snyder model for noncommutative spacetime (and its generalizations) is an ideal test ground to evaluate the physical relevance of different proposals for describing dimensionality in the quantum gravity regime. In fact, the (generalized) Snyder model is quite peculiar in having a highly non-trivial geometry, such that spacetime coordinates have a discrete spectrum, but it behaves as standard Minkowski spacetime once the physical constraint is imposed.

As a first investigation along this line, we compared the behavior of the thermal and spectral dimension associated to (generalized) Snyder spacetime. Quite interestingly, we found confirmation of earlier claims that the former is a good indicator of the presence of genuinely nontrivial physical properties of quantum spacetime models, or lack thereof. In fact, we found that the thermal dimension does not present any running and is constantly equal to the standard value of four.  On the other hand, analysis of the spectral dimension showed that its  behavior is dominated by off-shell properties of the model, so that for the Snyder spacetime it runs from the value of four in the infrared to zero in the ultraviolet. This confirms previous concerns regarding the physical relevance of such notion of dimensionality.

\section*{Acknowledgements}
F.G. acknowledges support from the John Templeton Foundation, Project 60609, ``Quantum Causal Structures'', from the research platform ``Testing Quantum and Gravity Interface with Single Photons'' (TURIS), and the Austrian Science Fund (FWF) through the project I-2526-N27 and the doctoral program ``Complex Quantum Systems'' (CoQuS) under Project W1210-N25. This publication was made possible through the support of a grant from the John Templeton Foundation. The opinions expressed in this publication are those of the authors and do not necessarily reflect the views of the John Templeton Foundation.

\bibliography{biblio}

\end{document}